\documentclass[11pt,oneside]{amsart}
\usepackage{geometry}
\usepackage{natbib}
\usepackage{caption,subcaption}
\usepackage{multicol,booktabs,colortbl,tabularx,graphicx,rotating,adjustbox,multicol,wrapfig}
\usepackage{amsmath,amssymb,amsthm,mathtools}
\def\T{{\mathrm{\scriptscriptstyle T}}}
\def\CD{{\mathrm{\scriptscriptstyle CD}}}

\usepackage{multicol,booktabs,colortbl,tabularx,graphicx,rotating,adjustbox,multicol,wrapfig}

\newtheorem{theorem}{Theorem}[section]

\newtheorem{lemma}[theorem]{Lemma}

\usepackage{algorithm}
\usepackage{algorithmic}
\begin{document}
%\begin{frontmatter}



\title{Compressed Covariance Estimation With Automated Dimension Learning}
\author{Gautam Sabnis, Debdeep Pati, Anirban Bhattacharya}


\address{}
\maketitle
\begin{abstract}
We propose a method for estimating a covariance matrix that can be represented as a sum of a low-rank matrix and a diagonal matrix. The proposed method compresses high-dimensional data, computes the sample covariance in the compressed space, and lifts it back to the ambient space via a decompression operation. A salient feature of our approach relative to existing literature on combining sparsity and low-rank structures in covariance matrix estimation is that we do not require the low-rank component to be sparse. A principled framework for estimating the compressed dimension using Stein's Unbiased Risk Estimation theory is demonstrated. Experimental simulation results demonstrate the efficacy and scalability of our proposed approach.

\end{abstract}

%\begin{keyword}



%\end{keyword}

%\end{frontmatter}

%%methods, including discriminant analysis \citep{friedman1989regularized,friedman2001elements}, principal component analysis \citep{zou2006sparse}%%
\section{Introduction}
Estimating a covariance matrix based on a sample of multivariate observations is a classical problem in statistics with applications across multitude of scientific disciplines. Time series analysis \citep{chen2013covariance,basu2015regularized}, portfolio optimization \citep{fan2008high,bai2011estimating}, gene networks \citep{butte2000discovering, schafer2005shrinkage}, climate studies \citep{houtekamer2001sequential, hamill2001distance, furrer2006covariance}, spatial data analysis \citep{kaufman2008covariance}, longitudinal data analysis \citep{smith2002parsimonious, wu2003nonparametric} among many other disciplines rely critically on the knowledge of the covariance structure. Recent efforts have focussed on high-dimensional data, where the dimension $p$ can be much larger than the sample size $n$ \citep{pourahmadi2011covariance,cai2016estimating,fan2016overview}. An unstructured $p \times p$ covariance matrix has $O(p^2)$ free entries. In moderate to high-dimensional situations, a general idea is to assume a restricted parameter space with much lower effective degrees of freedom. Examples include tapered covariance matrices \citep{furrer2007estimation}, bandable covariance matrices \citep{wu2003nonparametric,bickel2008regularized}, Toeplitz covariance matrices \citep{wu2009banding, mcmurry2010banded,xiao2012covariance}, sparse covariance matrices \citep{bickel2008covariance,karoui2008operator,karoui2008spectrum,rothman2009generalized,cai2011adaptive}, spiked sparse covariance matrices \citep{johnstone2001distribution, ma2013sparse, cai2015optimal}, covariances with a kronecker product structure \citep{werner2008estimation}, penalized likelihood estimation \citep{huang2006covariance,d2008first,lam2009sparsistency,ravikumar2011high} and regularization of principal components \citep{zou2006sparse,hoff2009hierarchical,johnstone2012consistency,cai2013sparse}. The challenge lies in the selection of appropriate structure or method for a specific data domain application. 

An alternative approach to controlling the complexity in covariance matrix estimation is through low intrinsic dimensionality. Low intrinsic dimensionality posits that the dependencies between the variables are captured by a small number of latent components, also called factors, explicitly seperating the common variation from variable-specific noise in the observed variables. Factor models in the high-dimensional regime have been used in a myriad of applications in economics and finance \citep{engle1981one, goldfarb2003robust}. There are two disparate literatures which express the covariance matrix as the sum of a low-rank and sparse matrix: (a) factor models \citep{bhattacharya2011sparse, fan2013large, pati2014posterior}, and (b) spiked covariance model \citep{johnstone2001distribution}. It is worthwhile to point out that the literature on both factor models and spiked covariance models assume the low-rank matrix is also sparse.

While the sparsity assumptions on the covariance matrix or the low rank component are well motivated for specific applications \citep{zou2006sparse, shen2008sparse, witten2009penalized,bhattacharya2011sparse}, recent studies reveal striking correlations between several genes/loci or gene networks and the features of the diseases they cause \citep{jimenez2001human, ideker2008protein}. The correlation between the attributes of complex disease genes is more extensive and stronger than previously thought. There is evidence that the etiology of many complex diseases involves, rather than a few genes/loci with large effects, many genes, each of which contributes a small risk, interacting with each other or with environmental risk factors to cause these complex diseases. These small effects are organized in networks/pathways that have distinct features. 

In this article, we attempt to mitigate the aforementioned gaps in the literature by (i) providing an efficient way of estimating covariance matrices which admit a decomposition of the form that is written as a sum of low-rank plus sparse matrix albeit with dense to moderately sparse low-rank structures, (ii) providing a concrete principled way of choosing the dimension of the low-rank matrix using Stein's Unbiased Risk Estimation (SURE) theory. We show via simulations that our approach is readily scalable to massive covariance matrices. The approach is based on a ``compression-decompression'' (C-D) mechanism. The C-D mechanism proceeds by projecting the high dimensional observations to a lower dimension to form compressed measurements and then decompress them back to the original dimension. The key idea is to use the sample covariance matrix of the compressed-decompressed data as an estimator of $\Sigma$ instead of the sample covariance matrix ${\widehat{\Sigma}}$. The covariance estimation problem considered in this paper is related to the covariance sketching problem considered in \cite{dasarathy2015sketching}. In covariance sketching, the goal is to estimate the covariance matrix of high dimensional random vectors based on the low-dimensional projections where the same instance of projection matrix is used across all observations. The choice of the projection matrix, in our setting, varies across observations and is integrated out to obtain the final estimator.

Our simulation studies demonstrate the efficacy of our approach and show the C-D estimator to be highly competitive to the recent state-of-the-art covariance estimators such as POET \citep{fan2013large} and adaptive thresholding \citep{cai2011adaptive}. 

\section{Compression - Decompression Covariance Estimator} \label{Estimator}
We now elucidate the compression-decompression (C-D) estimator. Let $X = [x_1, \ldots, x_n]$ denote a $p \times n$ data matrix with the columns $x_i$ independent and identically distributed from a $p$-variate distribution whose covariance we wish to estimate. We assume the data to be column-centered, so that $\mathbb{E}(x_i) = 0$ for all $i$, and thereby define $\widehat{\Sigma}_x = n^{-1} \sum_{i=1}^n x_i x_i^{\T} = XX^{\T}/n$ as the sample covariance matrix. The mechanism proceeds by projecting the data to a lower-dimensional space to form compressed measurements, computing the sample covariance in the compressed space, and lifting back to the ambient space via a decompression operation. 

Specifically, given $k < p$ and a $k \times p$ unitary matrix $\phi$ (with $\phi \phi^* = I_k$; $\phi^*$ denoting the complex conjugate of $\phi$), project the data $x_i$ from ${\mathbb{R}}^{p} \mapsto {\mathbb{R}}^{k}$ to create {\em{compressed data}} $w_i = \phi x_i$; let $W = [w_1, \ldots, w_n]$ denote the corresponding $k \times n$ matrix. The $k \times k$ sample covariance matrix 
$\Sigma_{w} = n^{-1} \sum_{i=1}^n w_i w_i^{*} = W W^{*}/n$ is expected to be more stable compared to the $p \times p$ matrix $\widehat{\Sigma}_x$. To obtain a $p \times p$ covariance estimate for our original problem, we decompress $\Sigma_{w}$ using the transformation $\phi^*$ to define
\begin{align}\label{eq:est}
\widehat{\Sigma}_{}(\phi; k) = \phi^* \Sigma_w \phi = \frac{1}{n} \phi^* (W W^{*}) \phi = \phi^*(\phi \widehat{\Sigma}_x \phi^*) \phi. 
\end{align} 
Observe that the regularization in this framework arises from the compression operation, which is entirely different from assuming $\ell_q$ type sparsity on the covariance matrix or its various decompositions. One of the motivations behind this approach comes from Theorem 8.1 of \citet{pati2014posterior}, where
it is shown that if the true $\Sigma_0 = \Lambda_0{\Lambda_0}^{\T} + \sigma^2 I_p$ with $\Lambda_0 \in {\mathbb{R}}^{p \times k}$, and one chooses $\phi = \Lambda_0$, then ${\widehat{\Sigma}}_{}$ concentrates around $\Sigma_0$ in operator norm with high probability when $k \ll p$ and $\sigma^2$ is {\em bounded}. 
It is important to mention here that a similar concentration for the sample covariance matrix in a spiked covariance model requires $\Sigma$ to be have a low effective rank \citep{bunea2015sample}, necessitating $\sigma^2 = O(1/p)$, which is fairly restrictive.

Evidently, the estimator in \eqref{eq:est} depends on two unknown parameters, the compression matrix $\phi$ and the projected dimension $k$. There has been recent work in the regression context \citep{guhaniyogi2013bayesian} where very high-dimensional covariates are projected to a lower-dimensional space using one particular instance of a random sensing matrix. However, this approach of fixing $\phi$ has very poor performance in our setting. An alternative way may be to estimate $\phi$ from the data is computationally intensive as it requires estimating $p \times k$ many parameters. Here, instead of trying to estimate the high-dimensional parameter $\phi$, we average over the ensemble of unitary matrices which can be performed in closed-form using random matrix results \citep{marzetta2011random}. Specifically, let $h(\cdot)$ denote the Haar measure on the space of $k \times p$ unitary matrices satisfying $h(\phi \Psi) = h(\phi)$ for all non-stochastic $p \times p$ unitary matrices $\Psi$. Letting $\mathbb{E}_{\phi}$ denote the expectation with respect to $h$, define the C-D estimator 
\begin{align*}
\widehat{\Sigma}_{\CD}(k) = \mathbb{E}_{\phi} \big[ \widehat{\Sigma}_{\CD}(\phi; k) \big].
\end{align*}
Based on recent random matrix techniques as in \citet{marzetta2011random}, the above expectation can be computed as
\begin{equation}\label{eq:shrinkage}
\widehat{\Sigma}_{\CD}(k) = \frac{k}{(p^2 - 1)p}\bigg[(pk - 1){{\widehat{\Sigma}}_{x}} + (p - k)\mbox{Tr}({\widehat{\Sigma}}_{x})I_{p}\bigg],
\end{equation}
where $\mbox{Tr}(A)$ denotes the trace of a matrix $A$. Clearly, averaging over the Haar distribution introduces an appropriate shrinkage on the sample covariance matrix resulting the estimator full rank even if $p \gg n$. As noted by  \citet{marzetta2011random}, \eqref{eq:shrinkage} bears resemblance with shrinkage estimators of the type $a \widehat{\Sigma}_{x}
+ (1-a) I_p$, where $0 < a < 1$ and $I_p$ is a $p \times p$ identity matrix, originally proposed by \citet{ledoit2004well}.  However, a key difference is that the effect of shrinkage $a \widehat{\Sigma}_{x}$ is compensated by $\mbox{Tr}(\widehat{\Sigma}_{x})I_{p}$ in \eqref{eq:shrinkage} instead of just $I_p$. This helps preserving the largest eigenvalues of the resulting estimator. 

A fundamental principle of statistical decision theory is that there
exists an interior optimum in the trade-off between bias and estimation error. One way of attaining this optimal trade-off is simply to take
a properly weighted average of the biased and unbiased estimators. Refer, for example, to the seminal work by  \citet{stein1956inadmissibility}, who showed that shrinking sample means towards a constant
can, under certain circumstances, improve accuracy. 
The crux of our method is to shrink the unbiased but very variable sample
covariance matrix towards the biased but less variable identity covariance matrix and
to thereby obtain a more efficient estimator. In addition, the resulting estimator, in \eqref{eq:shrinkage} is invertible
and well-conditioned, which is of crucial importance in settings where one needs to estimate the inverse, for example, in portfolio selection \citep{ledoit2003flexible}, Gaussian graphical models \citep{meinshausen2006high,yuan2007model} among others. 

\section{Choice of the compressed dimension} \label{dimension}
In this section, we propose a data-driven method for choosing the size of the compressed dimension. The main idea is to consider the Frobenius risk of the estimator for each choice of compressed dimension, $k$, and then choose that value which minimizes the Frobenius risk. Since the risk depends on the truth, we derive an unbiased estimator of the Frobenius risk curve and find the minimizer of the estimated risk curve to choose a value of $k$. We use Stein's Unbiased Risk Estimation (SURE) theory to find an unbiased estimator of the Frobenius risk associated with the C-D estimator. 

SURE was originally proposed in \cite{stein1981estimation} in deriving an unbiased estimator of the risk of James--Stein estimate. \citet{efron1986biased,efron2004estimation} applied SURE to prediction problems which was called {\em covariance penalty method}. \citet{li2016sure} proved the asymptotic properties of SURE information criterion for large bandable covariance matrices and proposed a family of generalized SURE ($\mbox{SURE}_{c}$) indexed by $c$ for covariance matrix estimation, where $c$ is some constant. For bandable covariance matrices, \citet{li2014asymptotic} claimed that $\mbox{SURE}_{2}$ and $\mbox{SURE}_{\log{}{n}}$ can be regarded as AIC and BIC analogues, respectively, for covariance matrix estimation problems. \cite{xiao2014theoretic} proposed an improved version of the banding estimator obtained in \cite{bickel2008regularized} and used SURE-type approach for selecting the bandwidth for the banding estimator. 

Let $R(k) = \mathbb{E}\|\widehat{\Sigma}_{\CD}(k) - {\Sigma}_0\|_{F}^{2}$ be the Frobenius risk associated with the proposed estimator for a fixed $k$. We use the following risk identity proved in \citet{yi2013sure},

\begin{equation}\label{eq:risk}
R(k) = \mathbb{E}\|\widehat{\Sigma}_{\CD}(k) - {\widehat{\Sigma}}_{x}\|_{F}^{2} - {\displaystyle\sum\limits_{i,j}^{p}}{{{\text{var}}({\widehat{\sigma}}_{ij})}} + 2{\displaystyle\sum\limits_{i,j}^{p}}{{{\text{cov}}({\widehat{\sigma}}_{ij}^{(k)}, {\widehat{\sigma}}_{ij})}}
\end{equation}
where ${\widehat{\Sigma}} = [{\widehat{\sigma}}_{ij}]$ is the usual sample covariance matrix, $\widetilde{\Sigma} = [{\widetilde{\sigma}}_{ij}]$ is the maximum likelihood estimator of the covariance matrix, with ${\widehat{\Sigma}} = \frac{n}{n-1}{\widetilde{\Sigma}}$, and $\widehat{\Sigma}_{\CD}(k) = [{\widehat{\sigma}}_{ij}^{(k)}]$ is the proposed covariance estimator \eqref{eq:shrinkage}. The third term on the right hand side of \eqref{eq:risk}  is referred to as the optimism \citep{efron2004estimation}. The second term on the right hand is the same for all estimators of $\Sigma_0$. 

Standard results from multivariate statistics \citep{anderson1984multivariate} imply 
${\text{var}}({\widehat{\sigma}}_{ij}) = \frac{\sigma_{ij}^{2} + \sigma_{ii}\sigma_{jj}}{n - 1}, {\text{var}}({\widehat{\sigma}}_{ii}) = \frac{2\sigma_{ii}^{2}}{n - 1}$ and ${\mbox{cov}}({\widehat{\sigma}}_{ll},{\widehat{\sigma}}_{ii}) = \frac{2 (n + 2)}{n (n - 1)}{\sigma_{il}}^{2} + \big(\frac{n+1}{n-1} - 1\big){\sigma_{ii}}{\sigma_{ll}}$. Lemma 3.1 provides unbiased estimators to each of these quantities which enables us to provide an unbiased estimator of $R(k)$. Recall ${\widehat{\Sigma}}$ and ${\widetilde{\Sigma}}$ defined in the previous paragraph. 
\begin{lemma} \label{Lemma A.2}
Let ${\widehat{\cdot}}$ denote an unbiased estimator of the quantity of interest. Then 
%\begin{equation} 
%{\mbox{var}}({\widehat{\sigma}}_{ij}) = \frac{\sigma_{ij}^{2} + \sigma_{ii}\sigma_{jj}}{n - 1}
%\end{equation} 
%and an estimate of ${\mbox{var}}({\widehat{\sigma}}_{ij})$ is given by 
\small\begin{align} 
{\widehat{\mbox{var}}}({\widehat{\sigma}}_{ij}) &= \frac{n^2(n^2 - n -4)}{{(n-1)}^{2}{(n^3 + n^2 - 2n - 4)}}{\widetilde{\sigma}}_{ij}^{2} + \frac{n^3}{(n-1)(n^3 + n^2 - 2n -4)}{\widetilde{\sigma}}_{ii}{\widetilde{\sigma}}_{jj}, \label{eq25} \\
{\widehat{\text{var}}}({\widehat{\sigma}}_{ii}) &= \frac{{n^2}\;{(2n^2 - 2n - 4)}}{{{(n-1)}^2}{(n^3 + n^2 - 2n - 4)}}\;{\widetilde{\sigma}}_{ii}^{2},  \label{eq27} \\
{\widehat{\mbox{cov}}}({\widehat{\sigma}}_{ll}, {\widehat{\sigma}}_{ii}) &= \frac{2{n^2}(n + 2)}{{(n - 1)}{(n^3 + n^2 - 2n - 4)}}\;{\widetilde{\sigma}}_{ij}^{2} + \frac{2{(n-2)}{n^2}}{{(n-1)}{(n^3 + n^2 - 2n - 4)}}{\widetilde{\sigma}}_{ii}{\widetilde{\sigma}}_{ll}. \label{eq29}
\end{align}
%\small\begin{equation} \label{eq25}
%{\widehat{\mbox{var}}}({\widehat{\sigma}}_{ij}) = \frac{n^2(n^2 - n -4)}{{(n-1)}^{2}{(n^3 + n^2 - 2n - 4)}}{\widetilde{\sigma}}_{ij}^{2} + \frac{n^3}{(n-1)(n^3 + n^2 - 2n -4)}{\widetilde{\sigma}}_{ii}{\widetilde{\sigma}}_{jj}
%\end{equation}
%%where $a_n = \frac{n^2(n^2 - n -4)}{{(n-1)}^{2}{(n^3 + n^2 - 2n - 4)}}$ and $b_n = \frac{n^3}{(n-1)(n^3 + n^2 - 2n -4)}$.  Also, 
%% \begin{equation} 
%%{\mbox{var}}({\widehat{\sigma}}_{ii}) = \dfrac{2\sigma_{ii}^{2}}{n - 1}
%%\end{equation} 
%%and an estimate of ${\mbox{var}}({\widehat{\sigma}}_{ii})$ is given by  
%\small\begin{equation} \label{eq27}
%{\widehat{\text{var}}}({\widehat{\sigma}}_{ii}) = \frac{{n^2}\;{(2n^2 - 2n - 4)}}{{{(n-1)}^2}{(n^3 + n^2 - 2n - 4)}}\;{\widetilde{\sigma}}_{ii}^{2}
%\end{equation}
%
%%\item \begin{equation}
%%{\mbox{cov}}({\widehat{\sigma}}_{ll}, {\widehat{\sigma}}_{ii}) = \frac{2 (n + 2)}{n (n - 1)}{\sigma_{il}}^{2} + \bigg(\frac{n+1}{n-1} - 1\bigg){\sigma_{ii}}{\sigma_{ll}}
%%\end{equation} 
%%and an estimate of ${\mbox{cov}}({\widehat{\sigma}}_{ll}, {\widehat{\sigma}}_{ii})$ is given by 
%\small\begin{equation} \label{eq29}
%{\widehat{\mbox{cov}}}({\widehat{\sigma}}_{ll}, {\widehat{\sigma}}_{ii}) = \frac{2{n^2}(n + 2)}{{(n - 1)}{(n^3 + n^2 - 2n - 4)}}\;{\widetilde{\sigma}}_{ij}^{2} + \frac{2{(n-2)}{n^2}}{{(n-1)}{(n^3 + n^2 - 2n - 4)}}{\widetilde{\sigma}}_{ii}{\widetilde{\sigma}}_{ll}
%\end{equation}
%%where $d_n = \frac{2{n^2}(n + 2)}{{(n - 1)}{(n^3 + n^2 - 2n - 4)}}$ and $e_n =  \frac{2{(n-2)}{n^2}}{{(n-1)}{(n^3 + n^2 - 2n - 4)}}$.
\end{lemma}
We now derive the SURE criterion for the Frobenius risk $R(k)$ defined in \eqref{eq:risk}.
\begin{theorem} \label{surethm}
An unbiased estimator of the Frobenius risk associated with $\widehat{\Sigma}_{\CD}(k)$ is given by 
\small
\begin{multline*} \label{eq17}
\mbox{SURE}(k) = (\eta - 1)^{2}\|{\widehat{\Sigma}} \circ {\widehat{\Sigma}} \|_{F} + p{\gamma^2} + 2{\gamma}(\eta - 1){\mbox{Tr}}({\widehat{\Sigma}}) \\
+ 2\bigg\{(a_n\eta + d_n\gamma)\bigg(\|{\widetilde{\Sigma}}\circ{\widetilde{\Sigma}}\|_{F} \;- \;\mbox{Tr}({\widetilde{\Sigma}}\circ{\widetilde{\Sigma}})\bigg) \notag \\ +  (b_n\eta +  c_n\gamma)\bigg({{\mbox{Tr}}({\widetilde{\Sigma}})}^{2} - {\mbox{Tr}}({\widetilde{\Sigma}} \circ {\widetilde{\Sigma}})\bigg) + c_n(\eta + \gamma){\mbox{Tr}}\bigg({\widetilde{\Sigma}} \circ {\widetilde{\Sigma}}\bigg)\bigg\}
\end{multline*}
\normalsize
where $\eta = \frac{k(pk - 1)}{p(p^2 - 1)}$, $\gamma = \frac{p - k}{p(p^2 - 1)}$, and $\Sigma_{1} \circ \Sigma_{2}$ denotes the Schur product between two matrices $\Sigma_1$ and $\Sigma_2$ of the same dimension. 
\end{theorem}

Having obtained the SURE criterion, we can now minimize it with respect to $k$ to select the compressed dimension. Specifically, define
\begin{equation} 
{\widehat{k}}^{\text{sure}} = {\mbox{argmin}}_k\; \mbox{SURE}(k). 
\end{equation}
Our simulation results show below that ${\widehat{k}^{\text{sure}}}$ provides an accurate estimate of the intrinsic dimensionality $k$ in the examples considered. 
\section{Experiments on Synthetic Data} \label{Experiments}
In this section, we consider a number of simulation cases to compare our proposed approach in terms of (a) accuracy of the SURE method in estimating $\widehat{k}^{\text{sure}}$, (b) accuracy of covariance matrix estimation in operator norm, and (c) accuracy of covariance matrix estimation in frobenius norm. The Frobenius norm ($\|\cdot\|_{F}$) and the operator norm ($\| \cdot \|_{2}$) are defined in the usual way with $\|A\|_{F} = \sqrt{\mbox{trace}(A^{\T}A)}$ and $\|A\|_{2} = s_{{max}}(A)$ where $s_{{max}}(A)$ denotes the largest singular value of $A$. 

We compare our method with Principal Orthognoal complement Thresholding (POET) of \citet{fan2013large} which is based on an additive decomposition of the covariance matrix in terms of a low rank matrix and a sparse residual covariance matrix. POET estimates the factors and the loadings by thresholding the principal components of the sample covariance matrix. We also compare with adaptive thresholding (AT) of \citet{cai2011adaptive} which thresholds the entries of the sample covariance matrix, with the resulting thresholded estimator ${\widehat{\Sigma}}$ being of the form ${\widehat{\Sigma}}_{jj'} = S_{jj'}1(|S_{jj'}| > {\delta_{\kappa_{jj'}}})$, where $\delta$ is the tuning parameter and $\kappa_{jj'}$ is a threshold specific to the corresponding entry of $S$. We choose the tuning parameter $\delta$ by 5-fold cross-validation as suggested by \citet{cai2011adaptive}. 

The two simulation settings considered here are described below: 

\begin{enumerate} 
\item $y_i$, $i = 1, \ldots, n$ are generated from $N_p(0,\Sigma_0)$, where $\Sigma_0 = \Lambda_0{\Lambda_0}^{\T} + \sigma_{0}^{2}{\mathrm{I}}_{p}$ and $\Lambda$ is a $p \times {\text{ktr}}$ matrix with $(1 - s) \times 100\%$ non-zero entries. Here, $s \in (0,1)$, is the {\textit{sparsity}} parameter. We choose different values of $s$ that lead to moderately sparse to dense covariance matrices. The nonzeros entries are independently drawn from a standard normal distribution. 

\item This setting is designed to illustrate the performance of our approach under model misspecification. We let $\Sigma = \Lambda_0\Lambda_0^{\T} + \Omega_{0}$, where $\Lambda_0$ is as in simulation setting (1), but $\Omega_0$ is nondiagonal corresponding to the covariance matrix of an autoregressive sequence with pure error variance 0.4 and autoregressive coefficient 0.1. 
\end{enumerate} 

For each simulation setting, we choose the sample size $n = 100$, dimension $p = 250, 500, 1000$ and the true number of factors ${\text{ktr}}$ = 10,50. For each $(n,p,\text{ktr})$ triplet, we consider 100 simulation replicates. We consider two different values of the sparsity parameter $s$, $s \in \{0.1,0.5\}$. $s = 0.1$ randomly sets $10\%$ entries in the factor loadings matrix to 0. This corresponds to the extreme situation in which there is not a single sparse entry in the true covariance matrix. Similarly, $s = 0.5$ randomly sets $50\%$ entries in the factor loadings matrix to 0 and corresponds to the moderately sparse regime.

To evaluate the accuracy of $\widehat{k}^{\text{sure}}$ , we compare with $k^{\text{opt}}$, defined as the minimizer of the true predictive risk function which assumes knowledge of the truth. For different settings, results across simulation replicates are summarized for two two simulation settings to compare the matrix norm differences between the estimator resulting from different methods and the truth. In particular, normalized average operator norm error ($\| \cdot \|_{2}/p$), in the top panel, and normalized average frobenius norm error ($\| \cdot \|_{F}/p$), in the bottom panel, across 100 replicates is provided with standard error in paranthesis. The tables indicate that the SURE method is very accurate in estimating $k^{\text{opt}}$. From Tables \ref{table1}, \ref{table2} it becomes evident as the number of model parameters increases and the sparsity reduces, the performance of AT and POET deteriorates, in terms of both operator and frobenius norms, due to the sparsity assumption on both estimators, while the C-D estimator has more robust performance. Even in Tables \ref{table3}, \ref{table4}, where the truth is misspecified for both C-D and AT, and in fact designed to favor POET, C-D performs better than its competitors. We also evaluate the performance of CD estimator across various levels of sparsity on the low-rank structure for fixed sample size $n$ and dimension $p$.  Figure \ref{spavserr} displays the superior performance of CD estimator in dense ($s = 0.1$) to moderately sparse ($s = 0.7$) regimes. 

\begin{table}[htbp!] 
\centering
\footnotesize
\caption{Simulation Setting 1 with $s = 0.5$. Top panel compares $\widehat{k}^{\text{sure}}$ with $k^{\text{opt}}$. Bottom two panels compare the proposed approach with POET and AT in terms of $\|{\widehat{\Sigma}} - {\Sigma_0}\|_{2}/p$ and $\|{\widehat{\Sigma}} - {\Sigma_0}\|_{F}/p$ respectively. Standard errors are in parenthesis.} \label{table1}
\begin{multicols}{7}
\centering
\begin{tabular}{|p{1.92cm}|p{1.92cm}|p{1.92cm}|p{1.92cm}|p{1.92cm}|p{1.92cm}|p{1.92cm}|} \toprule
{\bf ktr} & \multicolumn{3}{c}{{\bf{10}}}& \multicolumn{3}{c|}{{\bf{50}}} \\
\cmidrule(lr){2-4} \cmidrule(lr){5-7} \\
{\bf p} & \multicolumn{1}{c}{{\bf 250}} & \multicolumn{1}{c}{{\bf 500}} & \multicolumn{1}{c}{{\bf 1000}} & \multicolumn{1}{c}{{\bf 250}} & \multicolumn{1}{c}{{\bf 500}} & \multicolumn{1}{c|}{{\bf 1000}}\\
\hline\noalign{\smallskip}
${k^{\text{opt}}}$ & 240 & 480 & 950 & 210 & 420 & 830 \\  
${\widehat{k}}^{\text{sure}}$ & 240 & 480 & 950 & 210 & 410 & 820 \\  \hline \hline 
%${CS_{\mbox{opt}}}$ & ${\underset{(9.26)}{63.73}}$ & ${\underset{(16.56)}{138.29}}$ & ${\underset{(34.33)}{267.04}}$ & ${\underset{(7.67)}{158.33}}$ & ${\underset{(13.04)}{283.86}}$ & ${\underset{(17.22)}{544.20}}$ \\  
%CD & ${\underset{(9.26)}{63.73}}$ & ${\underset{(16.56)}{138.29}}$ & ${\underset{(34.33)}{267.04}}$ & ${\underset{(7.67)}{158.33}}$ & ${\underset{(12.19)}{285.61}}$ & ${\underset{(15.16)}{544.50}}$ \\  
CD & ${\underset{(0\cdot04)}{0\cdot25}}$ & ${\underset{(0\cdot03)}{0\cdot28}}$ & ${\underset{(0\cdot03)}{0\cdot27}}$ & ${\underset{(0\cdot03)}{0\cdot63}}$ & ${\underset{(0\cdot02)}{0\cdot57}}$ & ${\underset{(0\cdot02)}{0\cdot54}}$ \\  
%AT & ${\underset{(11.47)}{69.49}}$ & ${\underset{(25.56)}{148.07}}$ & ${\underset{(51.58)}{292.20}}$ & ${\underset{(22.05)}{219.81	}}$ & ${\underset{(39.67)}{385.46}}$ & ${\underset{(74.91)}{707.00}}$ \\ 
AT & ${\underset{(0\cdot05)}{0\cdot28}}$ & ${\underset{(0\cdot05)}{0\cdot30}}$ & ${\underset{(0\cdot05)}{0\cdot30}}$ & ${\underset{(0\cdot09)}{0\cdot88}}$ & ${\underset{(0\cdot08)}{0\cdot77}}$ & ${\underset{(0\cdot07)}{0\cdot71}}$ \\ 
%POET & ${\underset{(11.44)}{69.82}}$ & ${\underset{(25.50)}{149.20}}$ & ${\underset{(49.88)}{295.37}}$ & ${\underset{(22.76)}{233.47}}$ & ${\underset{(38.48)}{449.37}}$ & ${\underset{(72.36)}{919.83}}$ \\ 
POET & ${\underset{(0\cdot05)}{0\cdot28}}$ & ${\underset{(0\cdot05)}{0\cdot30}}$ & ${\underset{(0\cdot05)}{0\cdot30}}$ & ${\underset{(0\cdot09)}{0\cdot93}}$ & ${\underset{(0\cdot08)}{0\cdot90}}$ & ${\underset{(0\cdot07)}{0\cdot92}}$ \\ 
\hline
\hline
%${CS_{\mbox{opt}}}$ & ${\underset{(6.29)}{56.56}}$ & ${\underset{(13.66)}{107.07}}$ & ${\underset{(22.91)}{201.30}}$ & ${\underset{(7.23)}{123.94}}$ & ${\underset{(10.68)}{235.31}}$ & ${\underset{(13.71)}{428.57}}$ \\  
%CD & ${\underset{(6.29)}{56.56}}$ & ${\underset{(13.66)}{107.07}}$ & ${\underset{(26.69)}{201.78}}$ & ${\underset{(7.01)}{127.78}}$ & ${\underset{(10.68)}{235.31}}$ & ${\underset{(13.71)}{428.57}}$ \\  
CD & ${\underset{(0\cdot04)}{0\cdot56}}$ & ${\underset{(0\cdot03)}{0\cdot57}}$ & ${\underset{(0\cdot03)}{0\cdot54}}$ & ${\underset{(0\cdot03)}{2\cdot14}}$ & ${\underset{(0\cdot03)}{2\cdot14}}$ & ${\underset{(0\cdot03)}{2\cdot12}}$ \\  
%AT & ${\underset{(8.18)}{57.42}}$ & ${\underset{(16.51)}{112.83}}$ & ${\underset{(35.20)}{208.67}}$ & ${\underset{(14.25)}{157.07	}}$ & ${\underset{(29.32)}{306.54}}$ & ${\underset{(53.46)}{582.95}}$ \\ 
AT & ${\underset{(0\cdot06)}{0\cdot63}}$ & ${\underset{(0\cdot05)}{0\cdot63}}$ & ${\underset{(0\cdot05)}{0\cdot59}}$ & ${\underset{(0\cdot08)}{3\cdot50}}$ & ${\underset{(0\cdot08)}{3\cdot56}}$ & ${\underset{(0\cdot08)}{3\cdot57}}$ \\ 
%POET & ${\underset{(8.32)}{57.80}}$ & ${\underset{(16.41)}{112.59}}$ & ${\underset{(35.45)}{210.42}}$ & ${\underset{(13.62)}{158.76}}$ & ${\underset{(25.77)}{310.16}}$ & ${\underset{(48.4)}{588.35}}$ \\ 
POET & ${\underset{(0\cdot05)}{0\cdot60}}$ & ${\underset{(0\cdot05)}{0\cdot61}}$ & ${\underset{(0\cdot05)}{0\cdot58}}$ & ${\underset{(0\cdot08)}{2\cdot60}}$ & ${\underset{(0\cdot07)}{2\cdot62}}$ & ${\underset{(0\cdot08)}{2\cdot60}}$ \\ 
\hline
\end{tabular}
\end{multicols}
\end{table}
\begin{table}[htbp!]
\footnotesize
\caption{Simulation Setting 1 with $s = 0.1$. Top panel compares $\widehat{k}^{\text{sure}}$ with $k^{\text{opt}}$. Bottom two panels compare the proposed approach with POET and AT in terms of $\|{\widehat{\Sigma}} - {\Sigma_0}\|_{2}/p$ and $\|{\widehat{\Sigma}} - {\Sigma_0}\|_{F}/p$ respectively. Standard errors are in parenthesis.} \label{table2}
\begin{multicols}{7}
\centering
\begin{tabular}{|p{1.92cm}|p{1.92cm}|p{1.92cm}|p{1.92cm}|p{1.92cm}|p{1.92cm}|p{1.92cm}|} \toprule
{\bf ktr} & \multicolumn{3}{c}{{\bf{10}}}& \multicolumn{3}{c|}{{\bf{50}}} \\
\cmidrule(lr){2-4} \cmidrule(lr){5-7} \\
{\bf p} & \multicolumn{1}{c}{{\bf 250}} & \multicolumn{1}{c}{{\bf 500}} & \multicolumn{1}{c}{{\bf 1000}} & \multicolumn{1}{c}{{\bf 250}} & \multicolumn{1}{c}{{\bf 500}} & \multicolumn{1}{c|}{{\bf 1000}}\\
\hline
${k^{\text{opt}}}$ & 240 & 470 & 950 & 210 & 420 & 820 \\  
${\widehat{k}}^{\text{sure}}$ & 240 & 470 & 940 & 210 & 410 & 810 \\  \hline \hline 
%${CS_{\mbox{opt}}}$ & ${\underset{(9.26)}{63.73}}$ & ${\underset{(16.56)}{138.29}}$ & ${\underset{(34.33)}{267.04}}$ & ${\underset{(7.67)}{158.33}}$ & ${\underset{(13.04)}{283.86}}$ & ${\underset{(17.22)}{544.20}}$ \\  
%CD & ${\underset{(9.26)}{63.73}}$ & ${\underset{(16.56)}{138.29}}$ & ${\underset{(34.33)}{267.04}}$ & ${\underset{(7.67)}{158.33}}$ & ${\underset{(12.19)}{285.61}}$ & ${\underset{(15.16)}{544.50}}$ \\  
CD & ${\underset{(0\cdot06)}{0\cdot49}}$ & ${\underset{(0\cdot06)}{0\cdot52}}$ & ${\underset{(0\cdot06)}{0\cdot50}}$ & ${\underset{(0\cdot05)}{1\cdot16}}$ & ${\underset{(0\cdot04)}{1\cdot03}}$ & ${\underset{(0\cdot02)}{0\cdot95}}$ \\  
%AT & ${\underset{(11.47)}{69.49}}$ & ${\underset{(25.56)}{148.07}}$ & ${\underset{(51.58)}{292.20}}$ & ${\underset{(22.05)}{219.81	}}$ & ${\underset{(39.67)}{385.46}}$ & ${\underset{(74.91)}{707.00}}$ \\ 
AT & ${\underset{(0\cdot09)}{0\cdot52}}$ & ${\underset{(0\cdot10)}{0\cdot56}}$ & ${\underset{(0\cdot09)}{0\cdot56}}$ & ${\underset{(0\cdot12)}{1\cdot67}}$ & ${\underset{(0\cdot10)}{1\cdot39}}$ & ${\underset{(0\cdot06)}{1\cdot29}}$ \\ 
%POET & ${\underset{(11.44)}{69.82}}$ & ${\underset{(25.50)}{149.20}}$ & ${\underset{(49.88)}{295.37}}$ & ${\underset{(22.76)}{233.47}}$ & ${\underset{(38.48)}{449.37}}$ & ${\underset{(72.36)}{919.83}}$ \\ 
POET & ${\underset{(0\cdot09)}{0\cdot52}}$ & ${\underset{(0\cdot09)}{0\cdot56}}$ & ${\underset{(0\cdot09)}{0\cdot56}}$ & ${\underset{(0\cdot18)}{1\cdot74}}$ & ${\underset{(0\cdot12)}{1\cdot67}}$ & ${\underset{(0\cdot14)}{1\cdot61}}$ \\ 
\hline
\hline
%${CS_{\mbox{opt}}}$ & ${\underset{(6.29)}{56.56}}$ & ${\underset{(13.66)}{107.07}}$ & ${\underset{(22.91)}{201.30}}$ & ${\underset{(7.23)}{123.94}}$ & ${\underset{(10.68)}{235.31}}$ & ${\underset{(13.71)}{428.57}}$ \\  
%CD & ${\underset{(6.29)}{56.56}}$ & ${\underset{(13.66)}{107.07}}$ & ${\underset{(26.69)}{201.78}}$ & ${\underset{(7.01)}{127.78}}$ & ${\underset{(10.68)}{235.31}}$ & ${\underset{(13.71)}{428.57}}$ \\  
CD & ${\underset{(0\cdot07)}{0\cdot92}}$ & ${\underset{(0\cdot07)}{0\cdot98}}$ & ${\underset{(0\cdot07)}{0\cdot96}}$ & ${\underset{(0\cdot05)}{3\cdot88}}$ & ${\underset{(0\cdot06)}{3\cdot80}}$ & ${\underset{(0\cdot06)}{3\cdot76}}$ \\  
%AT & ${\underset{(8.18)}{57.42}}$ & ${\underset{(16.51)}{112.83}}$ & ${\underset{(35.20)}{208.67}}$ & ${\underset{(14.25)}{157.07	}}$ & ${\underset{(29.32)}{306.54}}$ & ${\underset{(53.46)}{582.95}}$ \\ 
AT & ${\underset{(0\cdot10)}{1\cdot01}}$ & ${\underset{(0\cdot10)}{1\cdot07}}$ & ${\underset{(0\cdot10)}{1\cdot05}}$ & ${\underset{(0\cdot42)}{6\cdot36}}$ & ${\underset{(0\cdot36)}{6\cdot30}}$ & ${\underset{(0\cdot17)}{6\cdot37}}$ \\ 
%POET & ${\underset{(8.32)}{57.80}}$ & ${\underset{(16.41)}{112.59}}$ & ${\underset{(35.45)}{210.42}}$ & ${\underset{(13.62)}{158.76}}$ & ${\underset{(25.77)}{310.16}}$ & ${\underset{(48.4)}{588.35}}$ \\ 
POET & ${\underset{(0\cdot08)}{0\cdot96}}$ & ${\underset{(0\cdot09)}{1\cdot03}}$ & ${\underset{(0\cdot09)}{1\cdot01}}$ & ${\underset{(0\cdot16)}{4\cdot68}}$ & ${\underset{(0\cdot14)}{4\cdot63}}$ & ${\underset{(0\cdot15)}{4\cdot59}}$ \\ 
\hline
\end{tabular}
\end{multicols}
\end{table}
\begin{table}[htbp!]
\footnotesize
\caption{Simulation Setting 2 with $s = 0.5$. Top panel compares $\widehat{k}^{\text{sure}}$ with $k^{\text{opt}}$. Bottom two panels compare the proposed approach with POET and AT in terms of $\|{\widehat{\Sigma}} - {\Sigma_0}\|_{2}/p$ and $\|{\widehat{\Sigma}} - {\Sigma_0}\|_{F}/p$ respectively. Standard errors are in parenthesis.} \label{table3}
\begin{multicols}{7}
\centering
\begin{tabular}{|p{1.92cm}|p{1.92cm}|p{1.92cm}|p{1.92cm}|p{1.92cm}|p{1.92cm}|p{1.92cm}|} \toprule
{\bf ktr} & \multicolumn{3}{c}{{\bf{10}}}& \multicolumn{3}{c|}{{\bf{50}}} \\
\cmidrule(lr){2-4} \cmidrule(lr){5-7} \\
{\bf p} & \multicolumn{1}{c}{{\bf 250}} & \multicolumn{1}{c}{{\bf 500}} & \multicolumn{1}{c}{{\bf 1000}} & \multicolumn{1}{c}{{\bf 250}} & \multicolumn{1}{c}{{\bf 500}} & \multicolumn{1}{c|}{{\bf 1000}}\\
\hline
${k^{\text{opt}}}$ & 240 & 480 & 950 & 210 & 420 & 830 \\  
${\widehat{k}}^{\text{sure}}$ & 240 & 480 & 940 & 210 & 410 & 810 \\  \hline \hline 
%${CS_{\mbox{opt}}}$ & ${\underset{(9.26)}{63.73}}$ & ${\underset{(16.56)}{138.29}}$ & ${\underset{(34.33)}{267.04}}$ & ${\underset{(7.67)}{158.33}}$ & ${\underset{(13.04)}{283.86}}$ & ${\underset{(17.22)}{544.20}}$ \\  
%CD & ${\underset{(9.26)}{63.73}}$ & ${\underset{(16.56)}{138.29}}$ & ${\underset{(34.33)}{267.04}}$ & ${\underset{(7.67)}{158.33}}$ & ${\underset{(12.19)}{285.61}}$ & ${\underset{(15.16)}{544.50}}$ \\  
CD & ${\underset{(0\cdot04)}{0\cdot31}}$ & ${\underset{(0\cdot03)}{0\cdot26}}$ & ${\underset{(0\cdot03)}{0\cdot28}}$ & ${\underset{(0\cdot03)}{0\cdot63}}$ & ${\underset{(0\cdot02)}{0\cdot57}}$ & ${\underset{(0\cdot02)}{0\cdot53}}$ \\  
%AT & ${\underset{(11.47)}{69.49}}$ & ${\underset{(25.56)}{148.07}}$ & ${\underset{(51.58)}{292.20}}$ & ${\underset{(22.05)}{219.81	}}$ & ${\underset{(39.67)}{385.46}}$ & ${\underset{(74.91)}{707.00}}$ \\ 
AT & ${\underset{(0\cdot06)}{0\cdot33}}$ & ${\underset{(0\cdot05)}{0\cdot29}}$ & ${\underset{(0\cdot05)}{0\cdot29}}$ & ${\underset{(0\cdot07)}{0\cdot88}}$ & ${\underset{(0\cdot05)}{0\cdot77}}$ & ${\underset{(0\cdot03)}{0\cdot71}}$ \\ 
%POET & ${\underset{(11.44)}{69.82}}$ & ${\underset{(25.50)}{149.20}}$ & ${\underset{(49.88)}{295.37}}$ & ${\underset{(22.76)}{233.47}}$ & ${\underset{(38.48)}{449.37}}$ & ${\underset{(72.36)}{919.83}}$ \\ 
POET & ${\underset{(0\cdot06)}{0\cdot33}}$ & ${\underset{(0\cdot05)}{0\cdot29}}$ & ${\underset{(0\cdot05)}{0\cdot30}}$ & ${\underset{(0\cdot08)}{0\cdot93}}$ & ${\underset{(0\cdot08)}{0\cdot92}}$ & ${\underset{(0\cdot07)}{0\cdot89}}$ \\ 
\hline
\hline
%${CS_{\mbox{opt}}}$ & ${\underset{(6.29)}{56.56}}$ & ${\underset{(13.66)}{107.07}}$ & ${\underset{(22.91)}{201.30}}$ & ${\underset{(7.23)}{123.94}}$ & ${\underset{(10.68)}{235.31}}$ & ${\underset{(13.71)}{428.57}}$ \\  
%CD & ${\underset{(6.29)}{56.56}}$ & ${\underset{(13.66)}{107.07}}$ & ${\underset{(26.69)}{201.78}}$ & ${\underset{(7.01)}{127.78}}$ & ${\underset{(10.68)}{235.31}}$ & ${\underset{(13.71)}{428.57}}$ \\  
CD & ${\underset{(0\cdot05)}{0\cdot57}}$ & ${\underset{(0\cdot04)}{0\cdot51}}$ & ${\underset{(0\cdot04)}{0\cdot53}}$ & ${\underset{(0\cdot03)}{2\cdot09}}$ & ${\underset{(0\cdot03)}{2\cdot09}}$ & ${\underset{(0\cdot03)}{2\cdot09}}$ \\  
%AT & ${\underset{(8.18)}{57.42}}$ & ${\underset{(16.51)}{112.83}}$ & ${\underset{(35.20)}{208.67}}$ & ${\underset{(14.25)}{157.07	}}$ & ${\underset{(29.32)}{306.54}}$ & ${\underset{(53.46)}{582.95}}$ \\ 
AT & ${\underset{(0\cdot06)}{0\cdot62}}$ & ${\underset{(0\cdot06)}{0\cdot56}}$ & ${\underset{(0\cdot06)}{0\cdot58}}$ & ${\underset{(0\cdot17)}{3\cdot39}}$ & ${\underset{(0\cdot20)}{3\cdot45}}$ & ${\underset{(0\cdot16)}{3\cdot49}}$ \\ 
%POET & ${\underset{(8.32)}{57.80}}$ & ${\underset{(16.41)}{112.59}}$ & ${\underset{(35.45)}{210.42}}$ & ${\underset{(13.62)}{158.76}}$ & ${\underset{(25.77)}{310.16}}$ & ${\underset{(48.4)}{588.35}}$ \\ 
POET & ${\underset{(0\cdot06)}{0\cdot60}}$ & ${\underset{(0\cdot05)}{0\cdot53}}$ & ${\underset{(0\cdot05)}{0\cdot55}}$ & ${\underset{(0\cdot05)}{2\cdot52}}$ & ${\underset{(0\cdot08)}{2\cdot54}}$ & ${\underset{(0\cdot07)}{2\cdot55}}$ \\ 
\hline
\end{tabular}
\end{multicols}
\end{table}

\begin{table}[htbp!]
\footnotesize
\caption{Simulation Setting 2 with $s = 0.1$. Top panel compares $\widehat{k}^{\text{sure}}$ with $k^{\text{opt}}$. Bottom two panels compare the proposed approach with POET and AT in terms of $\|{\widehat{\Sigma}} - {\Sigma_0}\|_{2}/p$ and $\|{\widehat{\Sigma}} - {\Sigma_0}\|_{F}/p$ respectively. Standard errors are in parenthesis.} \label{table4} 
\begin{multicols}{7}
\centering
\begin{tabular}{|p{1.92cm}|p{1.92cm}|p{1.92cm}|p{1.92cm}|p{1.92cm}|p{1.92cm}|p{1.92cm}|} \toprule
{\bf ktr} & \multicolumn{3}{c}{{\bf{10}}}& \multicolumn{3}{c|}{{\bf{50}}} \\
\cmidrule(lr){2-4} \cmidrule(lr){5-7} \\
{\bf p} & \multicolumn{1}{c}{{\bf 250}} & \multicolumn{1}{c}{{\bf 500}} & \multicolumn{1}{c}{{\bf 1000}} & \multicolumn{1}{c}{{\bf 250}} & \multicolumn{1}{c}{{\bf 500}} & \multicolumn{1}{c|}{{\bf 1000}}\\
\hline
${k^{\text{opt}}}$ & 240 & 480 & 950 & 210 & 420 & 830 \\  
${\widehat{k}}^{\text{sure}}$ & 240 & 470 & 940 & 210 & 410 & 810 \\  \hline \hline 
%${CS_{\mbox{opt}}}$ & ${\underset{(9.26)}{63.73}}$ & ${\underset{(16.56)}{138.29}}$ & ${\underset{(34.33)}{267.04}}$ & ${\underset{(7.67)}{158.33}}$ & ${\underset{(13.04)}{283.86}}$ & ${\underset{(17.22)}{544.20}}$ \\  
%CD & ${\underset{(9.26)}{63.73}}$ & ${\underset{(16.56)}{138.29}}$ & ${\underset{(34.33)}{267.04}}$ & ${\underset{(7.67)}{158.33}}$ & ${\underset{(12.19)}{285.61}}$ & ${\underset{(15.16)}{544.50}}$ \\  
CD & ${\underset{(0\cdot06)}{0\cdot49}}$ & ${\underset{(0\cdot06)}{0\cdot50}}$ & ${\underset{(0\cdot05)}{0\cdot50}}$ & ${\underset{(0\cdot05)}{1\cdot17}}$ & ${\underset{(0\cdot04)}{1\cdot04}}$ & ${\underset{(0\cdot02)}{0\cdot95}}$ \\  
%AT & ${\underset{(11.47)}{69.49}}$ & ${\underset{(25.56)}{148.07}}$ & ${\underset{(51.58)}{292.20}}$ & ${\underset{(22.05)}{219.81	}}$ & ${\underset{(39.67)}{385.46}}$ & ${\underset{(74.91)}{707.00}}$ \\ 
AT & ${\underset{(0\cdot09)}{0\cdot54}}$ & ${\underset{(0\cdot10)}{0\cdot53}}$ & ${\underset{(0\cdot09)}{0\cdot56}}$ & ${\underset{(0\cdot17)}{1\cdot67}}$ & ${\underset{(0\cdot10)}{1\cdot43}}$ & ${\underset{(0\cdot09)}{1\cdot26}}$ \\ 
%POET & ${\underset{(11.44)}{69.82}}$ & ${\underset{(25.50)}{149.20}}$ & ${\underset{(49.88)}{295.37}}$ & ${\underset{(22.76)}{233.47}}$ & ${\underset{(38.48)}{449.37}}$ & ${\underset{(72.36)}{919.83}}$ \\ 
POET & ${\underset{(0\cdot09)}{0\cdot54}}$ & ${\underset{(0\cdot10)}{0\cdot53}}$ & ${\underset{(0\cdot09)}{0\cdot55}}$ & ${\underset{(0\cdot18)}{1\cdot75}}$ & ${\underset{(0\cdot13)}{1\cdot66}}$ & ${\underset{(0\cdot12)}{1\cdot62}}$ \\ 
\hline
\hline
%${CS_{\mbox{opt}}}$ & ${\underset{(6.29)}{56.56}}$ & ${\underset{(13.66)}{107.07}}$ & ${\underset{(22.91)}{201.30}}$ & ${\underset{(7.23)}{123.94}}$ & ${\underset{(10.68)}{235.31}}$ & ${\underset{(13.71)}{428.57}}$ \\  
%CD & ${\underset{(6.29)}{56.56}}$ & ${\underset{(13.66)}{107.07}}$ & ${\underset{(26.69)}{201.78}}$ & ${\underset{(7.01)}{127.78}}$ & ${\underset{(10.68)}{235.31}}$ & ${\underset{(13.71)}{428.57}}$ \\  
CD & ${\underset{(0\cdot08)}{0\cdot90}}$ & ${\underset{(0\cdot09)}{0\cdot92}}$ & ${\underset{(0\cdot08)}{0\cdot94}}$ & ${\underset{(0\cdot03)}{3\cdot85}}$ & ${\underset{(0\cdot06)}{3\cdot81}}$ & ${\underset{(0\cdot05)}{3\cdot75}}$ \\  
%AT & ${\underset{(8.18)}{57.42}}$ & ${\underset{(16.51)}{112.83}}$ & ${\underset{(35.20)}{208.67}}$ & ${\underset{(14.25)}{157.07	}}$ & ${\underset{(29.32)}{306.54}}$ & ${\underset{(53.46)}{582.95}}$ \\ 
AT & ${\underset{(0\cdot11)}{0\cdot98}}$ & ${\underset{(0\cdot12)}{0\cdot99}}$ & ${\underset{(0\cdot11)}{1\cdot03}}$ & ${\underset{(0\cdot59)}{6\cdot18}}$ & ${\underset{(0\cdot36)}{6\cdot36}}$ & ${\underset{(0\cdot28)}{6\cdot33}}$ \\ 
%POET & ${\underset{(8.32)}{57.80}}$ & ${\underset{(16.41)}{112.59}}$ & ${\underset{(35.45)}{210.42}}$ & ${\underset{(13.62)}{158.76}}$ & ${\underset{(25.77)}{310.16}}$ & ${\underset{(48.4)}{588.35}}$ \\ 
POET & ${\underset{(0\cdot09)}{0\cdot94}}$ & ${\underset{(0\cdot11)}{0\cdot95}}$ & ${\underset{(0\cdot10)}{0\cdot98}}$ & ${\underset{(0\cdot17)}{4\cdot64}}$ & ${\underset{(0\cdot13)}{4\cdot64}}$ & ${\underset{(0\cdot13)}{4\cdot57}}$ \\ 
\hline
\end{tabular}
\end{multicols}
\end{table}
%\begin{figure}[htbp!]
%\centering
%\begin{subfigure}{.45\textwidth}
%\centering
%\includegraphics[width = 1.2\linewidth]{frobvsn.jpg} 
%\caption{Fixed $p$, varying $n$} 
%\label{fig:sub1}
%\end{subfigure} 
%\begin{subfigure}{.45\textwidth} 
%\centering
%\includegraphics[width = 1.2\linewidth]{frobvsp.jpg} 
%\caption{Fixed $n$, varying $p$} 
%\label{fig:sub2}
%\end{subfigure} 
%\caption{Frobenius norm error comparisons for the three estimators in two different settings.} 
%\label{fig:test} 
%\end{figure}
%\begin{figure}[htbp!]
%\centering
%\begin{subfigure}{.45\textwidth}
%\centering
%\includegraphics[width = 1.1\linewidth]{operatorvsn.jpg} 
%\caption{Fixed $p$, varying $n$} 
%\label{fig:sub1}
%\end{subfigure} 
%\begin{subfigure}{.45\textwidth} 
%\centering
%\includegraphics[width = 1.1\linewidth]{operatorvsp.jpg} 
%\caption{Fixed $n$, varying $p$} 
%\label{fig:sub2}
%\end{subfigure} 
%\caption{Operator norm error comparisons for the three estimators in two different settings.} 
%\label{fig:test} 
%\end{figure}

\section{Discussion}
\begin{figure}[htp!]
\begin{center}$
\begin{array}{cc}
\includegraphics[width=70mm]{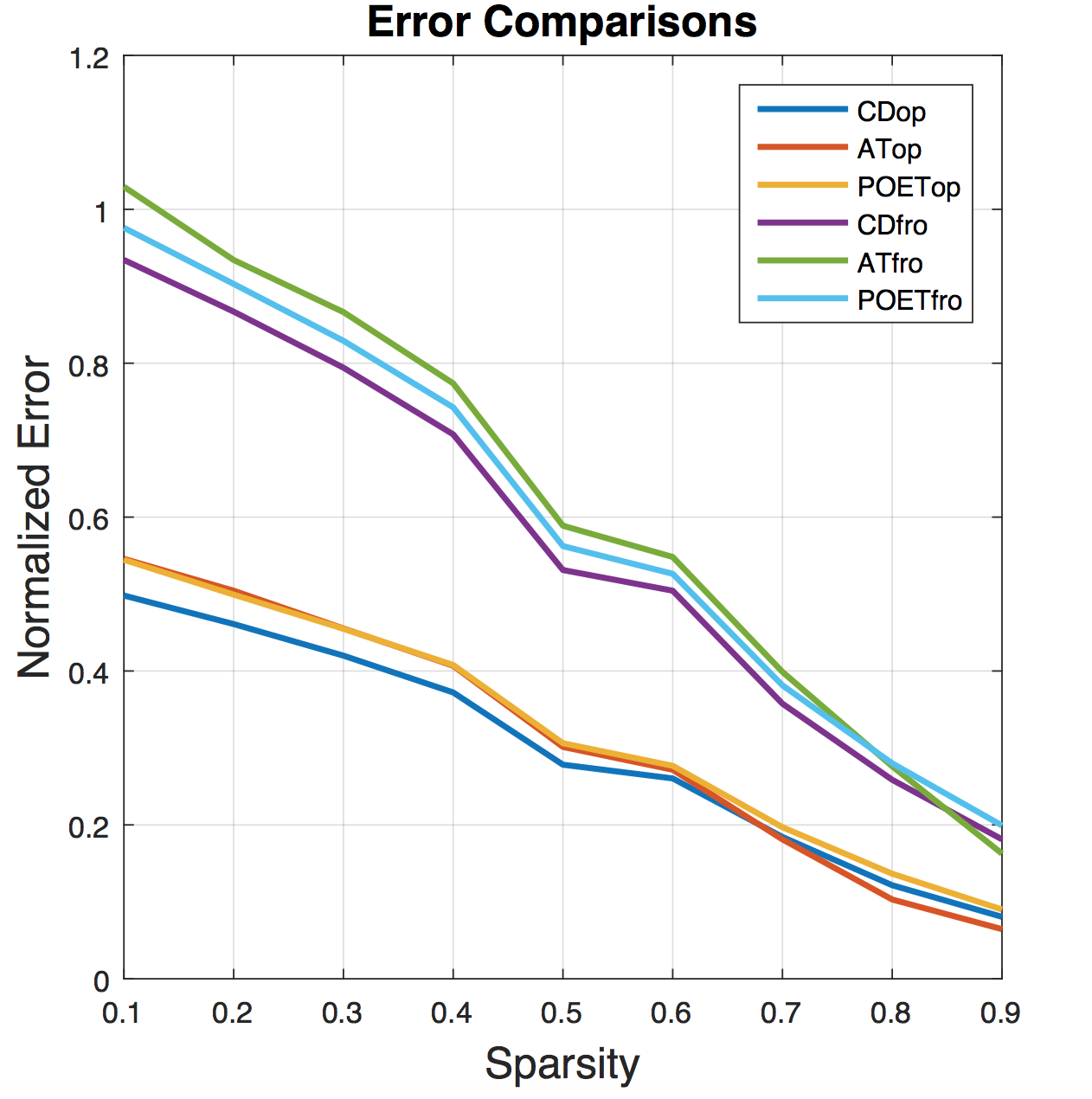}&
\includegraphics[width=70mm]{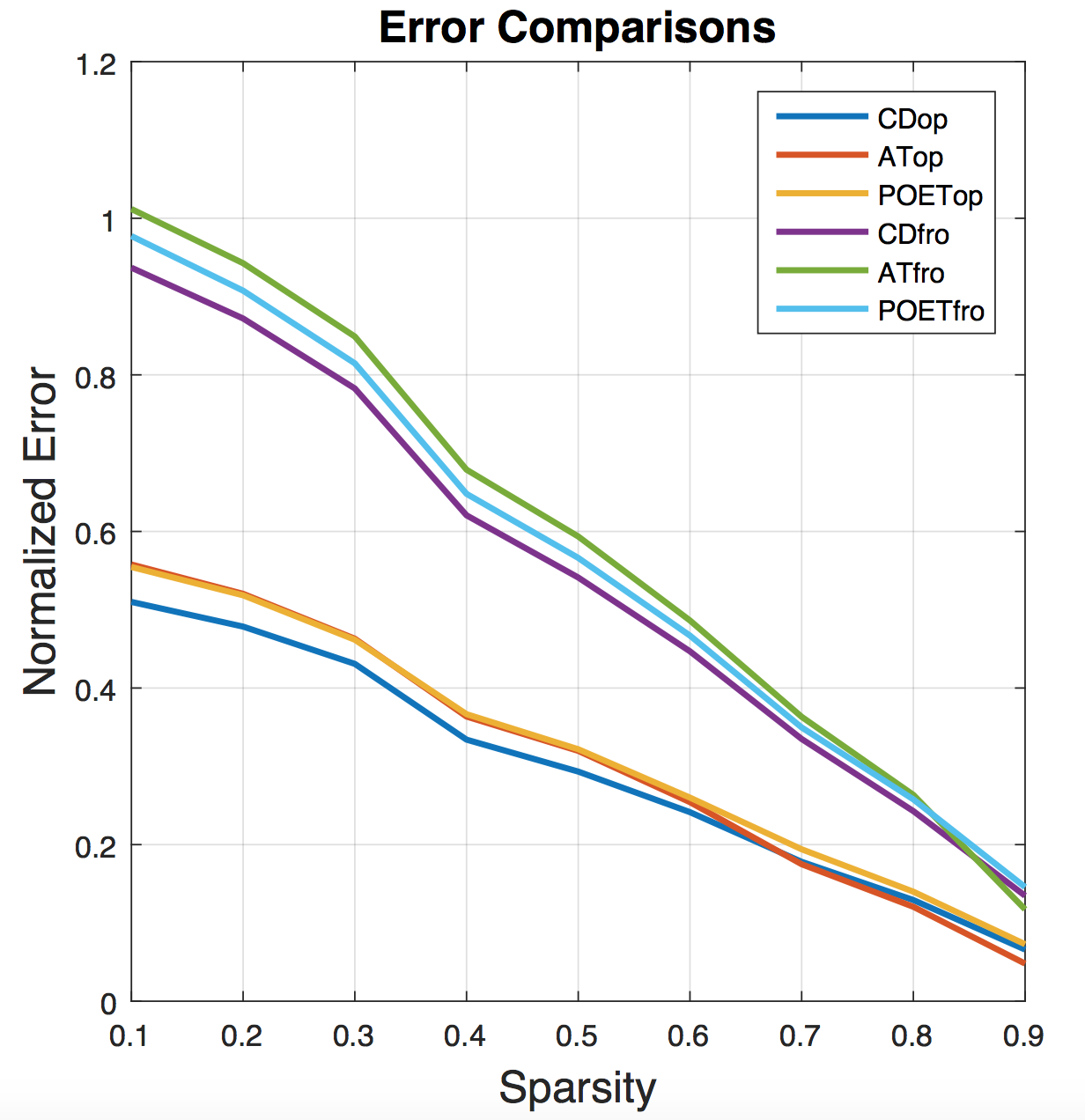}\\
\text{Simulation Setting 1} & \text{Simulation Setting 2}
\end{array}$
\end{center}
\caption{Operator(op) and Frobenius(fro) norm error comparisons averaged over 100 replicates for different  levels of sparsity for $n = 100$ and $p = 250$. } 
\label{spavserr} 
\end{figure}

In this article, we developed a simple but useful method for estimating covariance matrices with dense low-rank structures under the assumption of low intrinsic dimensionality. We also provide a principled framework for choosing the size of the low- rank dimension using SURE theory. We observe excellent performances of the proposed method in terms of scalability to high dimensions and capability of dealing with model misspecification. From Figure \ref{spavserr}, CD estimator outperforms AT and POET in dense $(s = 0.1)$ to moderately sparse $(s =  0.7$) regimes. 
%\begin{figure}[h]
%\centering 
%\includegraphics[width = 100mm]{sim1spavserr}
%\end{figure}
%
%\begin{figure}[h]
%\centering 
%\includegraphics[width = 100mm]{sim2spavserr}
%\end{figure}
%\begin{center}
%\begin{figure}[htbp!] 
%\begin{subfigure}{.50\textwidth}
%\includegraphics[width = 1.7\linewidth]{sim1spavserr} 
%\caption{Simulation Setting 1} 
%\label{fig:sub1}
%\end{subfigure} 
%\begin{subfigure}{.50\textwidth} 
%\includegraphics[width = 1.7\linewidth]{sim2spavserr} 
%\caption{Simulation Setting 2} 
%\label{fig:sub2}
%\end{subfigure} 
%\caption{Operator(op) and Frobenius(fro) norm error comparisons averaged over 100 replicates for different  levels of sparsity for $n = 100$ and $p = 250$. } 
%\label{spavserr} 
%\end{figure}
%\end{center}

%\begin{figure}[htbp!] 
%\includegraphics[width = 1.0\linewidth]{sim1spavserr} 
%\includegraphics[width = 1.0\linewidth]{sim2spavserr} \\
%Setting 1 \quad Setting 2
%\label{fig:sub2}
%\caption{Operator(op) and Frobenius(fro) norm error comparisons averaged over 100 replicates for different  levels of sparsity for $n = 100$ and $p = 250$. } 
%\label{spavserr} 
%\end{figure}

\newpage \section*{References} 
\bibliographystyle{elsarticle-harv}
\bibliography{referencesup}

\section*{Appendix}
\appendix
%\begin{lemma}
%Let $\widehat{\Sigma}_{\CD}(k) = [\sigma_{ij}^{(k)}]$ dentoe the C-D estimator. Then the unbiased estimator of the Frobenius risk associated with $\widehat{\Sigma}_{\CD}(k)$ is given by 
%\small
%\begin{eqnarray} \label{eq17}
%\mbox{SURE}(k) = (\eta - 1)^{2}\|{\widehat{\Sigma}} \circ {\widehat{\Sigma}} \|_{F} + p{\gamma^2} + 2{\gamma}(\eta - 1){\mbox{Tr}}({\widehat{\Sigma}}) + \notag \\
%2\bigg\{[a_n\eta + d_n\gamma]\bigg(\|{\widetilde{\Sigma}}\circ{\widetilde{\Sigma}}\|_{F} \;- \;\mbox{Tr}({\widetilde{\Sigma}}\circ{\widetilde{\Sigma}})\bigg) \notag \\ +  [b_n\eta +  c_n\gamma]\bigg({{\mbox{Tr}}({\widetilde{\Sigma}})}^{2} - {\mbox{Tr}}({\widetilde{\Sigma}} \circ {\widetilde{\Sigma}})\bigg) + c_n(\eta + \gamma)\notag \\{\mbox{Tr}}\bigg({\widetilde{\Sigma}} \circ {\widetilde{\Sigma}}\bigg)\bigg\}
%\end{eqnarray}
%\normalsize
%where $\eta = \frac{k(pk - 1)}{p(p^2 - 1)}$, $\gamma = \frac{p - k}{p(p^2 - 1)}$ and $\Sigma_{1} \circ \Sigma_{2}$ denotes the Schur product between two matrices $\Sigma_1$ and $\Sigma_2$ of the same dimension. 
%\end{lemma}

\section{Proof of Lemma \ref{Lemma A.2}}
We obtain unbiased estimators of \eqref{eq25},\eqref{eq27}, and \eqref{eq29}. 
Suppose $\{X_i\}_{i=1}^{n}$ is a random sample from $N(\mu, \Sigma)$ where, without loss of generality, we let $\mu = 0$. We have, 
\begin{align} \label{A.1}
{\mathbb{E}}(({\widetilde{\sigma}}_{ij}^{s})^{2})  = \frac{n}{n-1}{\sigma_{ij}^{2}} + \frac{\sigma_{ii}\sigma_{jj}}{n-1} 
\end{align}
and, 
\begin{align} \label{A.2}
{\mathbb{E}}({\widetilde{\sigma}}_{ii}^{s}{\widetilde{\sigma}}_{jj}^{s}) = \frac{n+1}{n-1}{\sigma_{ii}\sigma_{jj}} + \frac{2(n+2)}{n(n-1)}{\sigma_{ij}^{2}}. 
\end{align} 
\eqref{A.1} and \eqref{A.2} are obtained from \citet{yi2013sure}. Solving \eqref{A.1} and \eqref{A.2} simultaneously, we obtain unbiased estimators for ${\sigma_{ij}^{2}}$ and ${\sigma_{ii}\sigma_{jj}}$ below
\small \begin{align} \label{A.3}
{\mathbb{E}}\bigg[\frac{n(n^2 - 1)}{n^3 + n^2 - 2n -4}({\widetilde{\sigma}}_{ij}^{s})^{2} - \frac{n(n - 1)}{n^3 + n^2 - 2n -4}{\widetilde{\sigma}}_{ii}^{s}{\widetilde{\sigma}}_{jj}^{s}\bigg] = {\sigma_{ij}^{2}}. 
\end{align} 
\small\begin{align} \label{A.4}
{\mathbb{E}}\bigg[\frac{2(n-1)(n+2)}{2n + 4 - n^3 - n^2}({\widetilde{\sigma}}_{ij}^{s})^{2} - \frac{n^2(n - 1)}{2n + 4 - n^3 - n^2}{\widetilde{\sigma}}_{ii}^{s}{\widetilde{\sigma}}_{jj}^{s}\bigg] =  {\sigma_{ii}}{\sigma_{jj}}.
\end{align} 
An unbiased estimator of ${\mbox{Var}}({\widetilde{\sigma}}_{ij}^{s})$ is given by 
\begin{equation} \label{A.5}
\widehat{\mbox{Var}}({\widetilde{\sigma}}_{ij}^{s}) = \frac{{\widehat{\sigma}}_{ij}^{2} + {\widehat{\sigma_{ii}\sigma_{jj}}}}{n - 1}. 
\end{equation} 
Substituting \eqref{A.3} and \eqref{A.4} in \eqref{A.5} gives \eqref{eq25}.

(\ref{eq27}) is obtained from (\ref{eq25}) trivially. To obtain (\ref{eq29}), note that 
\begin{equation} \label{A.6}
{\mbox{Cov}}({\widetilde{\sigma}}_{jj}^{s}, {\widetilde{\sigma}}_{ii}^{s}) = \frac{2(n + 2)}{n(n - 1)}{\sigma_{ij}^{2}} + \frac{2}{n - 1}{\sigma_{ii}}{\sigma_{jj}}.
\end{equation} 
Substituting the unbiased estimators of ${\sigma}_{ij}^{2}$ and ${\sigma_{ii}}{\sigma_{jj}}$ in \eqref{A.6}, obtained from \eqref{A.3} and \eqref{A.4}, gives us (\ref{eq29}). This completes the proof. 

\section{Proof of Theorem \ref{surethm}}
We analyze each term in \eqref{eq:risk} one at a time. Consider $\|\widehat{\Sigma}_{\CD}(k)  - {\widehat{\Sigma}}\|_{F}^{2}$, a natural unbiased estimator of ${\mathbb{E}}\|\widehat{\Sigma}_{\CD}(k)  - {\widehat{\Sigma}}\|_{F}^{2}$. Then 
\small\begin{align} \label{B.1}
\|\widehat{\Sigma}_{\CD}(k)  - {\widehat{\Sigma}}\|_{F}^{2} & = \displaystyle\sum\limits_{i,j}^{p}{\bigg[(\eta - 1){\widehat{\sigma}}_{ij} + {\gamma}I(i = j)\bigg]}^{2}  \notag \\
& = {\displaystyle\sum\limits_{i \ne j}^{p}}{(\eta - 1)^{2}{\widehat{\sigma}}_{ij}^{2}} + \displaystyle\sum\limits_{i=j}^{p}\bigg[(\eta - 1){\widehat{\sigma}}_{ii} + \gamma\bigg]^{2} \notag \\ 
& = (\eta - 1)^{2}\|{\widehat{\Sigma}} \circ {\widehat{\Sigma}} \|_{F} + p{\gamma^2} + 2{\gamma}(\eta - 1){\mbox{Tr}}({\widehat{\Sigma}}),
\end{align}
where \eqref{B.1} is obtained by noting that ${\displaystyle\sum\limits_{i\ne j}^{p}}{\widetilde{\sigma}}_{ij}^{2} = {\displaystyle\sum\limits_{i=1}^{p}}{\displaystyle\sum\limits_{\underset{j \ne i}{j = 1}}^{p}}{\widetilde{\sigma}}_{ij}^{2} = \|{\widetilde{\Sigma}}\circ{\widetilde{\Sigma}}\|_{F} \;- \;\mbox{Tr}({\widetilde{\Sigma}}\circ{\widetilde{\Sigma}})$. Consider optimism, the third term on the right hand side of \eqref{eq:risk}. Then 
\begin{align} \label{B.2}
{\text{optimism}} & = {\displaystyle\sum\limits_{i \ne j}^{p}} {\eta}{\mbox{Var}}({\widehat{\sigma}}_{ij}) + {\displaystyle\sum\limits_{i = 1}^{p}}\bigg\{\eta{\mbox{Var}}({\widehat{\sigma}}_{ii}) + \gamma\;{\mbox{cov}}\bigg({\displaystyle\sum\limits_{\underset{l \ne i}{l = 1}}^{p}}{\widehat{\sigma}}_{ll} + {\widehat{\sigma}}_{ii}, {\widehat{\sigma}}_{ii}\bigg)\bigg\} \notag \\
& = {\displaystyle\sum\limits_{i \ne j}^{p}} {\eta}{\mbox{Var}}({\widehat{\sigma}}_{ij}) + {\displaystyle\sum\limits_{i = 1}^{p}}\bigg\{(\eta + \gamma){\mbox{Var}}({\widehat{\sigma}}_{ii}) + \gamma{\displaystyle\sum\limits_{\underset{l \ne i}{l = 1}}^{p}}{\mbox{cov}}\bigg({\widehat{\sigma}}_{ll}, {\widehat{\sigma}}_{ii}\bigg)\bigg\}.
\end{align}
Using Lemma \ref{Lemma A.2}, we have 
\begin{align} \label{B.3}
{\widehat{\text{optimism}}} & = {\eta}a_n{\displaystyle\sum\limits_{i\ne j}^{p}}{\widetilde{\sigma}}_{ij}^{2} + {\eta}b_n{\displaystyle\sum\limits_{i=1}^{p}}{\widetilde{\sigma}}_{ii}{\displaystyle\sum\limits_{j \ne i}^{p}}{\widetilde{\sigma}}_{jj} + c_n(\gamma + \eta){\displaystyle\sum\limits_{i = 1}^{p}}{\widetilde{\sigma}}_{ii}^{2} + d_n{\gamma}{\displaystyle\sum\limits_{i \ne l}^{p}}{\widetilde{\sigma}}_{il}^{2} + e_n{\displaystyle\sum\limits_{i \ne l}^{p}}{\widetilde{\sigma}}_{ii}{\widetilde{\sigma}}_{ll} \notag \\
& = 2\bigg\{(a_n\eta + d_n\gamma)\bigg(\|{\widetilde{\Sigma}}\circ{\widetilde{\Sigma}}\|_{F} \;- \;\mbox{Tr}({\widetilde{\Sigma}}\circ{\widetilde{\Sigma}})\bigg) +  (b_n\eta +  c_n\gamma) \notag \\
& \bigg({{\mbox{Tr}}({\widetilde{\Sigma}})}^{2} - {\mbox{Tr}}({\widetilde{\Sigma}} \circ {\widetilde{\Sigma}})\bigg) + c_n(\eta + \gamma){\mbox{Tr}}\bigg({\widetilde{\Sigma}} \circ {\widetilde{\Sigma}}\bigg)\bigg\},
\end{align}
where \eqref{B.3} is obtained by writing ${\displaystyle\sum\limits_{i=1}^{p}}{\widetilde{\sigma}}_{ii}{\displaystyle\sum\limits_{\underset{j \ne i}{j = 1}}^{p}}{\widetilde{\sigma}}_{jj} = {{\mbox{Tr}}({\widetilde{\Sigma}})}^{2} - {\mbox{Tr}}({\widetilde{\Sigma}} \circ {\widetilde{\Sigma}})$ and ${\displaystyle\sum\limits_{i=1}^{p}}{\widetilde{\sigma}}_{ii}^{2} = {\mbox{Tr}}({\widetilde{\Sigma}} \circ {\widetilde{\Sigma}})$.

The proof is completed by combining \eqref{B.1} and \eqref{B.3} to obtain an unbiased estimator of the Frobenius risk of ${\widehat{\Sigma}}_{CD}(k)$. 

\end{document}